\documentclass[a4paper]{mem}
\usepackage{natbib}
\usepackage{graphicx}
\usepackage[a4paper]{hyperref}
\idline{73}{23}
\begin{document}
   \title{The Pisa Evolutionary Library}
   \author{
          S. Degl'Innocenti \inst{1,2},
          P. Cariulo \inst{1},          
          P.G. Prada Moroni \inst{1,2,3}
          \and
          M. Marconi \inst{4}}

\mail{scilla@df.unipi.it}

   \institute{Dipartimento di Fisica, Universit\`a di Pisa, Via Buonarroti 2, 56127 Pisa, Italy 
              \and  INFN, Sezione di Pisa,Via Buonarroti 2, 56127 Pisa, Italy 
		\and Osservatorio Astronomico di Collurania, via Maggini, 64100
       Teramo, Italy
              \and  Osservatorio Astronomico di Capodimonte, via Moiariello 16, 80131 Napoli, Italy
             }
\abstract{We supplement recent evolutionary computations 
of canonical stellar models (i.e. with inefficient core overshooting)
with Z=0.0002, 0.0004, 0.0006, 0.001, 0.004, 0.008 and
suitable assumptions about the original He content. 
Evolutionary results have been 
compared with observational data in order to properly calibrate the
 efficiency of the surface convection. 
On this basis, we follow the evolution of stellar models
in the mass range 0.6 to 11 M$_{\odot}$ from the Main Sequence (MS) until
 C ignition or the onset of thermal pulses in the advanced
Asymptotic Giant Branch (AGB) phase, presenting cluster isochrones covering
the range of ages from 20 Myr to 20 Gyr.  To allow a comparison with
evolutionary investigations appeared in the recent literature, we
computed  additional sets of models which take into
account moderate core overshooting during the H burning phase
 (for the metallicities suitable for stars in the Magellanic Clouds: Z=0.004, Z=0.008).
 Selected predictions constraining the
cluster ages are discussed, presenting a calibration of the difference
in magnitude between the luminous MS termination and the He burning
giants in terms of the cluster age. Both evolutionary tracks and
isochrones have been made available at the URL http://gipsy.cjb.net
in the ``Pisa Evolutionary Library'' directory.
} 
\authorrunning{Degl'Innocenti et al.}
\titlerunning{Pisa Evolutionary Library} 
\maketitle
\section{Model calibrations}
There is the need for an extensive
 grid of stellar models with a range of suitable chemical compositions and ages
  to analyze the history of stellar populations in the Milky Way
 and in nearby galaxies. 
Recent theoretical computations all rely on the assumption of an
efficient overshooting in the convective cores \cite{yi01,girardi00,salasnich00}.
 The parallel availability of canonical evolutionary
\begin{figure*}
\centering
\resizebox{\hsize}{!}{\includegraphics[clip=true]{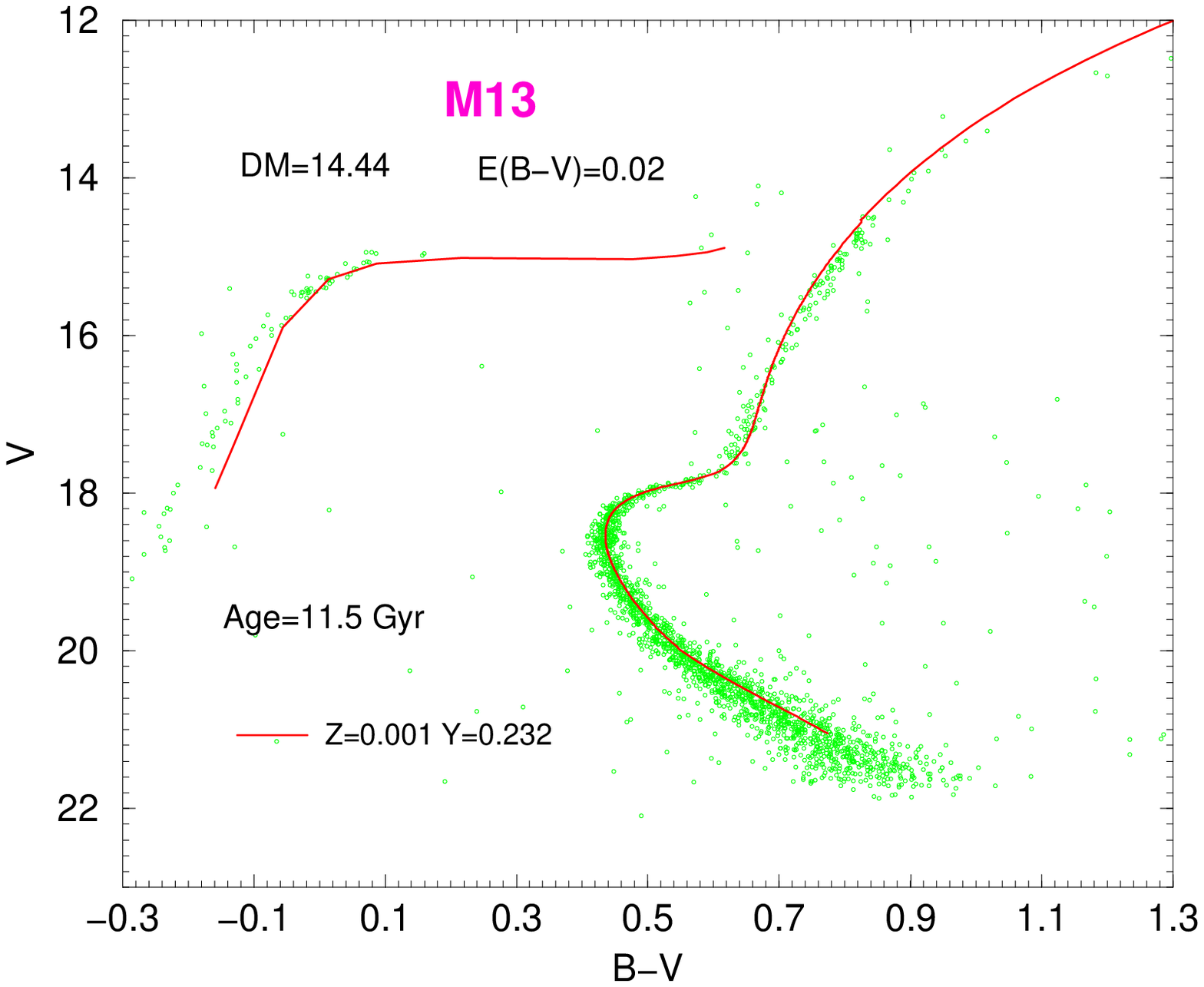}
   \includegraphics[]{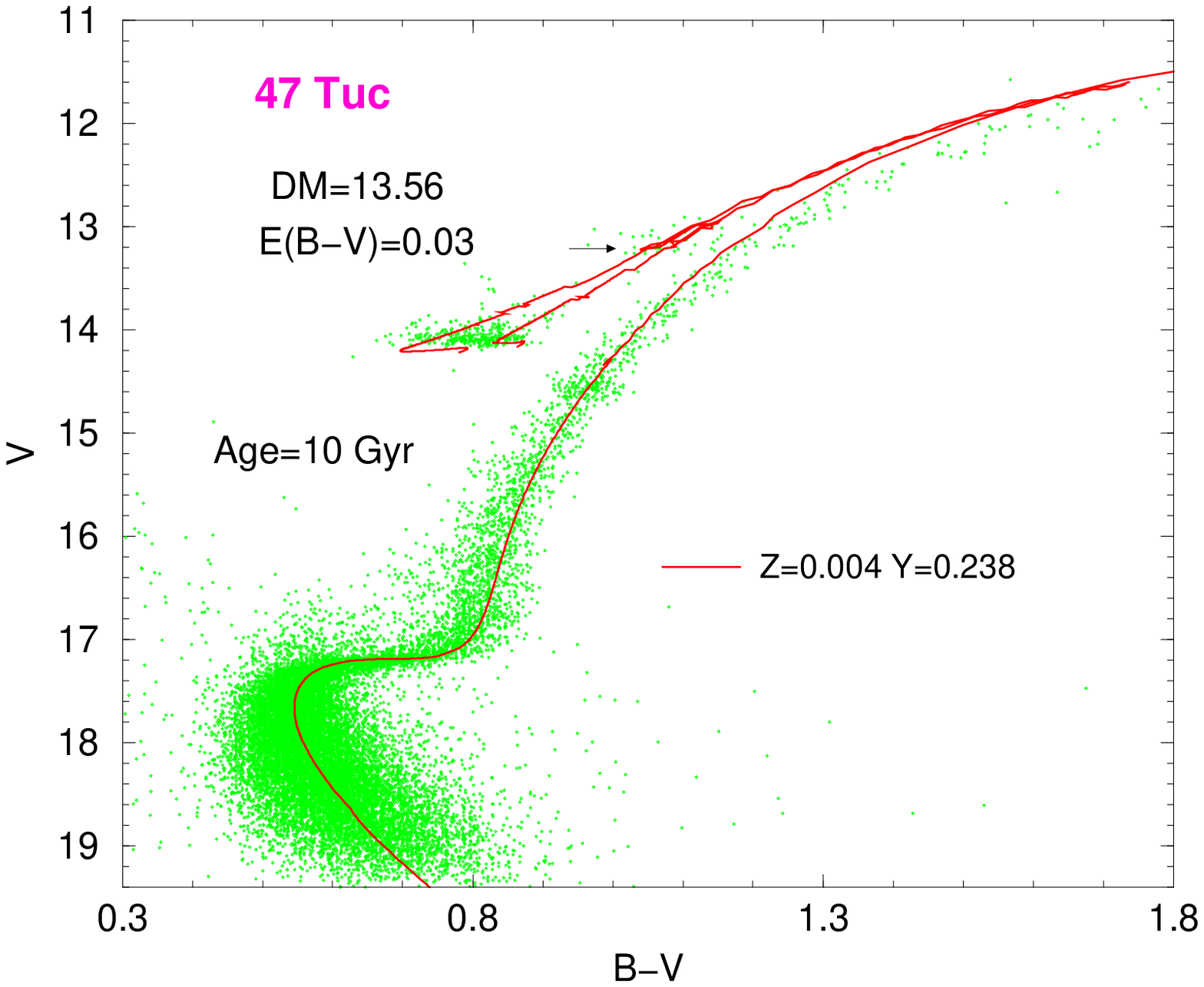}}
\caption{The observed (V,B-V) CM diagrams of M13 ([Fe/H]=-1.39$\pm$0.07)
 and 47 Tuc ([Fe/H]=-0.70$\pm$0.03) and
 the isochrones best fits as obtained
 by adopting $\alpha$=1.9 for Z=0.004 Y=0.238 and
 $\alpha$=2.3 for Z=0.001 Y=0.232. For 47 Tuc two horizontal branch models are also shown,
together with the predicted Asymptotic Giant Barnch (AGB) bottom luminosity (arrow). 
Metallicity values from \citealp{carrettagratton}.} 
\label{f1} 
\end{figure*}
%
   \begin{figure*}
   \centering
   \resizebox{\hsize}{!}{\includegraphics[clip=true]{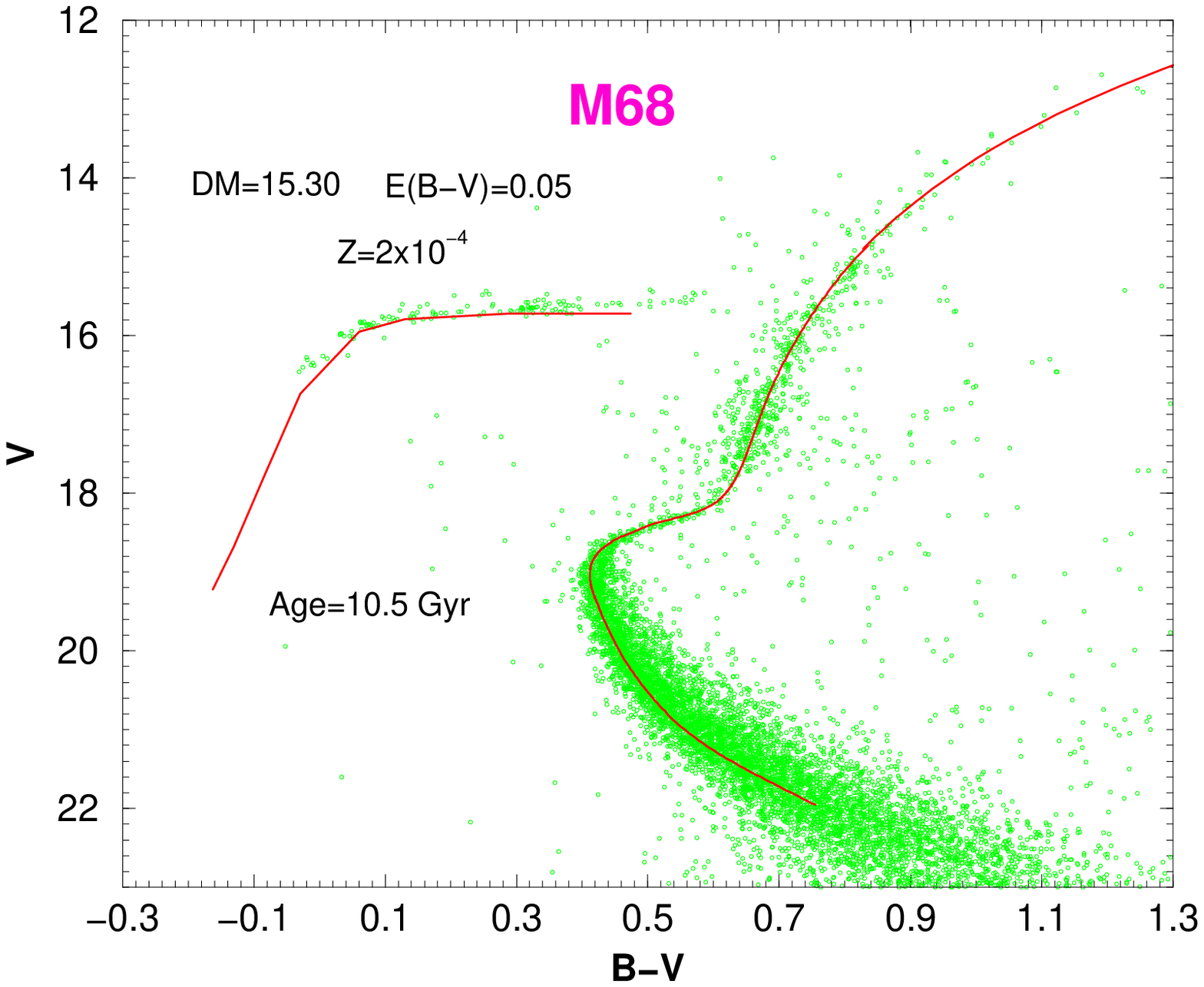}
   \includegraphics[]{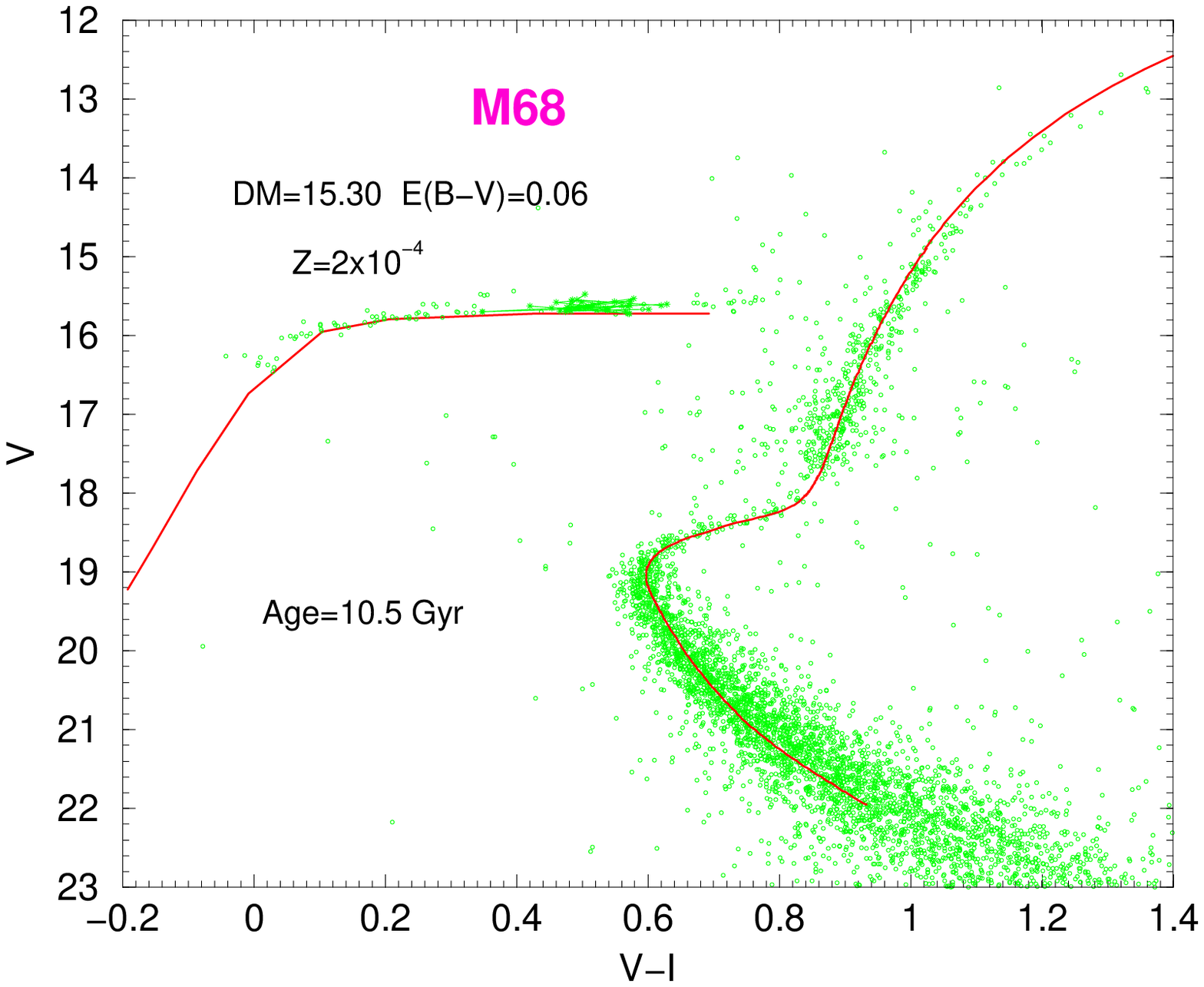}}
\caption{The  observed CM diagram of M68, with superimposed
the best fit isochrone in the B-V (left panel) and V-I (right panel)
 colors by adopting $\alpha$=2.0.} 
\label{f2} 
\end{figure*}
%
results, employing the Schwarzschild criterion are needed.
   \begin{figure*}
   \centering
   \resizebox{\hsize}{!}{\includegraphics[clip=true]{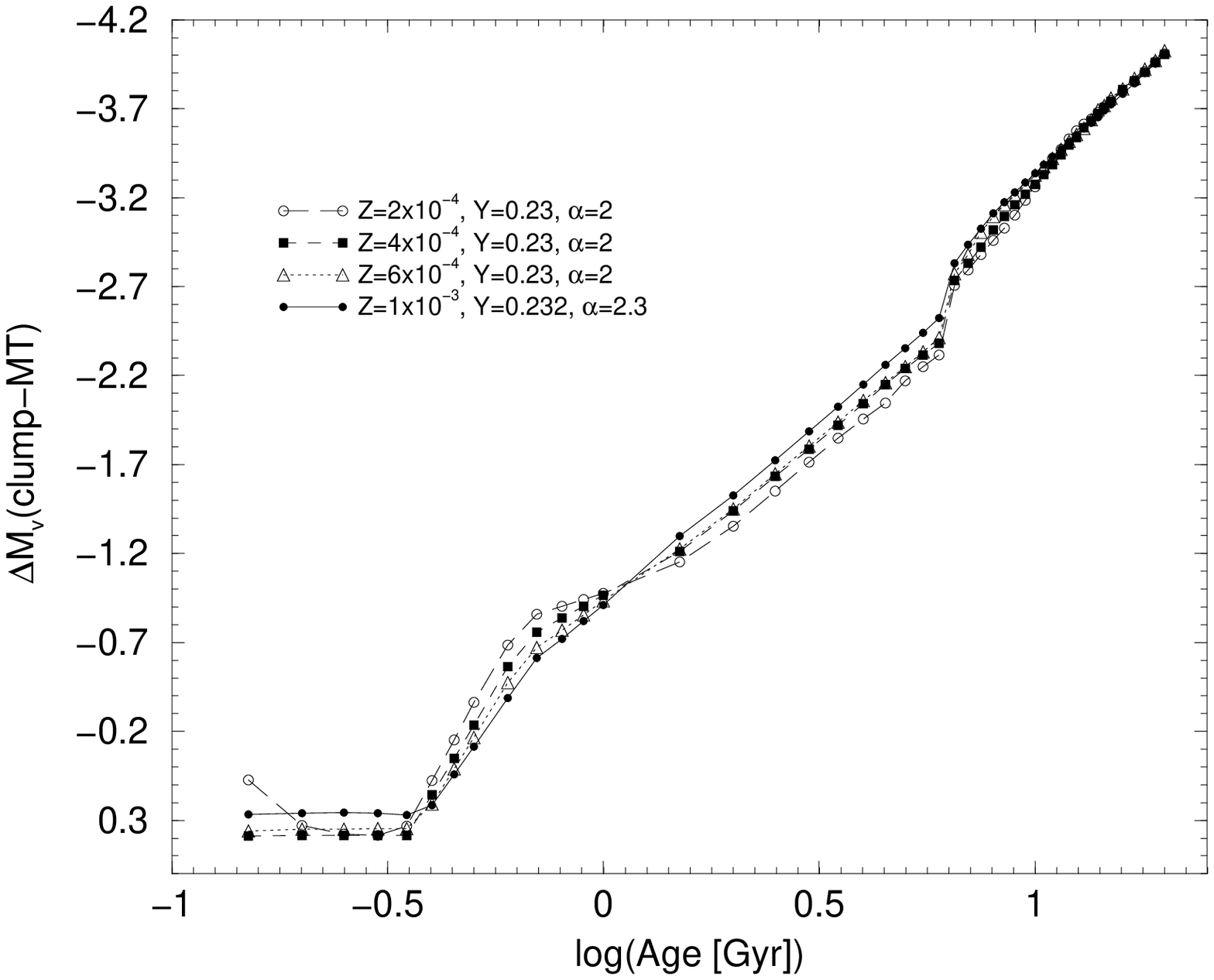}
   \includegraphics[]{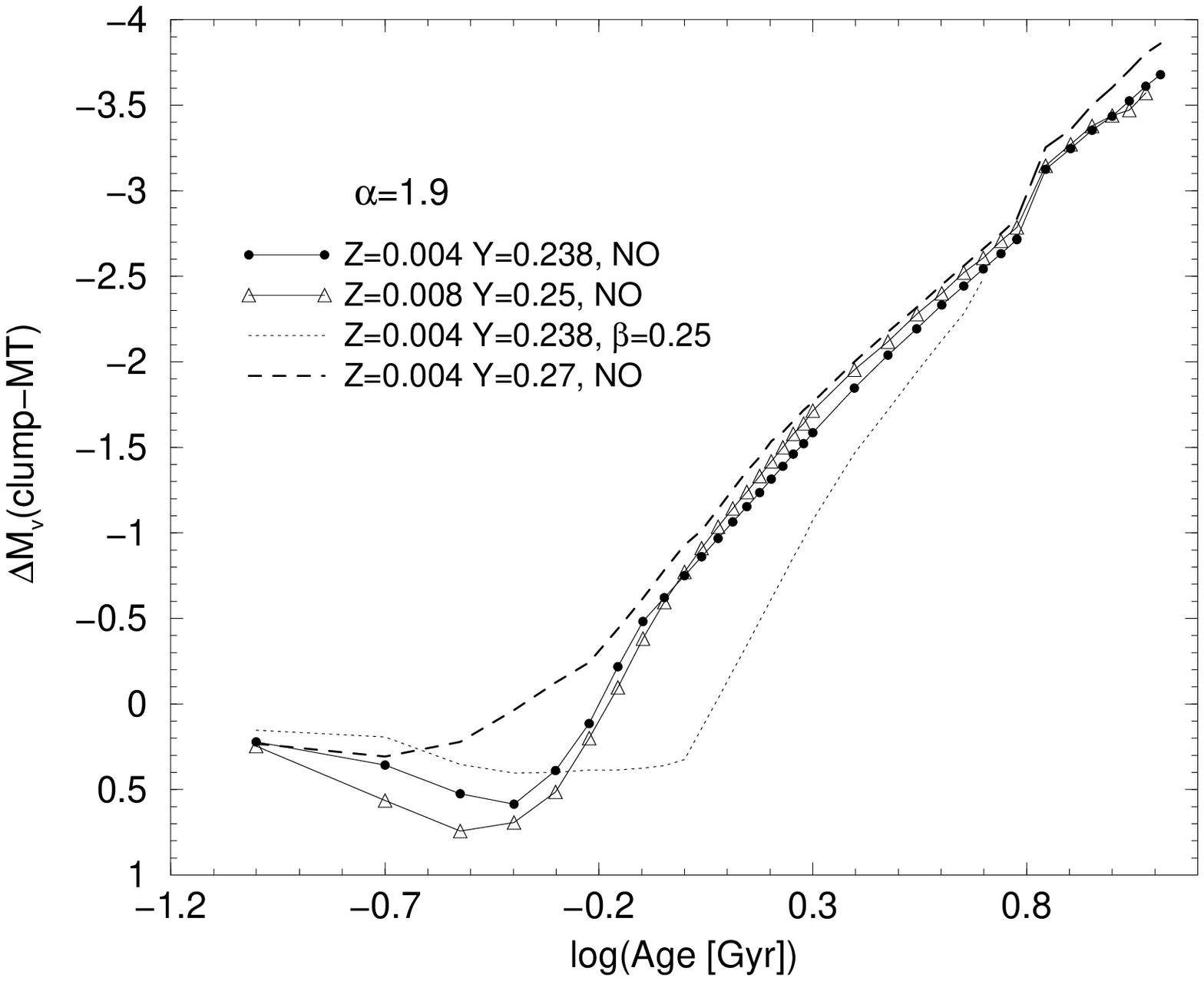}}
\caption{
$\Delta$M$_V$ $\mathrm {^{TO}_{clump}}$ as a function of the
logarithm of the age for different chemical compositions and core 
overshooting efficiency (NO= canonical models). For ages larger than 5 Gyr 
convective cores vanish. For ages higher than 6 Gyr the classical 
definition of TO and the HB visual magnitude at the RR Lyrae
 region have been adopted.
} 
\label{f3} 
\end{figure*}
%
Our theoretical approach has been already satisfactorily
 tested vis-a-vis Solar Standard Models \cite{scilla}
 and young metal rich galactic clusters with Hipparcos
 parallaxes \cite{castellani01,castellani02}. 
Here we extended this procedure to lower metallicities. 
 Original helium abundances have been obtained by assuming 
a primordial helium content Y=0.23 and $\Delta$Y/$\Delta$Z=2.5
 \cite{pagel98}.
The efficiency of the superadiabatic convection at the upper envelope
 of stellar structures, specified by the mixing length parameter,
 has been calibrated against the observations.
As an example, Figures \ref{f1} and \ref{f2} show the comparison of our
 isochrones with well observed, not too reddened, globular clusters
 with appropriate [Fe/H]. All the clusters appear well fitted, with distance
moduli, reddenings and ages well within the range of values accepted
in the recent literature \cite{sarajedini91,gratton97,rood99,carretta00}.
Moreover Fig. \ref{f2} shows the encouraging evidence that the best fit
obtained in the (V,B-V) plane also holds when
passing to V-I, supporting the consistency of the
adopted model atmospheres by \citealp{castelli99}. The physical inputs
adopted in our code have been described in \citealp{castellani03}
 (Paper I), \citealp{ciacio97,cassisi98}.
 In the present canonical models the convective 
boundaries are fixed by the Schwarzschild criterion; however we computed 
additional sets of models (Z=0.004 Y=0.238, Z=0.008 Y=0.25, see paper I)
by allowing an efficient core overshooting in the H burning phase, as modelled
according to \citealp{castellani00} with ${l}_{\rm ov}=\beta{\rm
H}_{\rm p}$, and $\beta$ values covering the range suggested in the
recent literature \cite{girardi00}.  Comparisons with
the \citealp{girardi00} and \citealp{yi01} models can be found in
Paper I, in \citealp{cariulo} and in \citealp{castellani00}.
Detailed tables for both tracks and isochrones are available at the
URL http://gipsy.cjb.net in the ``Pisa evolutionary library''
directory including also all the adopted chemical compositions,
tables with the distribution of Zero Age Horizontal Branch (ZAHB)
models together with tables reporting the Horizontal Branch (HB) and
AGB evolution of all these models. To allow a detailed comparison among
different evolutionary computations, the web site contains
 files listing the evolution of structural characteristics
 of three selected models (M=0.9, 2.0 and 4.0 M$_{\odot}$)
 for all chemical compositions.
\section{An age indicator}
Isochrone fitting is the best way for dating stellar clusters.
However, the availability of more direct and simple ``age
indicators" has been proved to be useful.
Since the pioneering paper by \citealp{iben},
 for old globulars, the so called ``vertical method'',
 based on the difference between the Turn-Off and the helium burning
HB phase has been widely used \cite{stetson96,cassisi98,cassisi99}.
 A similar method can also be used for younger clusters,
 calibrating the difference in magnitude between the bright
 terminal MS (MT) and the clump of He burning giants
 evaluated, respectively, at the brightest magnitude reached just 
after the overall contraction  (H exhaustion) phase and the minimum 
luminosity of the He clump region \cite{salaris02,castellani99,udalski98}.
 We used our new isochrones to calibrate these
 parameters in terms of the cluster ages.
 The aim is to derive an easy-to-use observational parameter that is able
to provide a correct order of magnitude for the age
 from the color-magnitude diagram of a cluster, independently
 of its distance.  
Figure \ref{f3} shows the run of $\Delta$M$_{\rm V}$ as a function of the cluster
age for the selected chemical compositions. As already known 
(see, e.g., Fig. 11 in \citealp{ccs}) $\Delta$M$_{\rm V}$
works as an univocal indicator only for ages larger than 700-800
Myr, whereas for lower ages there is
some ambiguity. Data in the same figure shows 
the dependence of the calibration on the assumptions about 
overshooting and He content.
In papers I and II one can find tables which give, for each value of the
cluster age, the V magnitude of the MT and the difference
in visual magnitude between HB and MT.
 For each age, the original mass of stars populating the
He burning clump is also reported. Data in these tables can be
useful in several ways. For example, by looking at the CM
diagram of cluster NGC2420 \cite{at90} from the
observed difference in magnitude between the He-clump and the MS
termination ($\Delta$M$_{\rm V}\sim$ 1.5 mag) one finds for the cluster
a (canonical) age of the order of 2 Gyr, in agreement with the
results obtained through the fit of the CMD diagram \cite{pols98,prada}
 but without the need for a complicate isochrone fitting procedure. 
\begin{acknowledgements}
We warmly thank Vittorio Castellani for his advice and Steve Shore for
 a careful reading of the manuscript.
\end{acknowledgements}
\bibliographystyle{aa}

\end{document}